\documentclass{elsart}
\usepackage{ifpdf}
\usepackage{graphicx,natbib,amssymb,lineno,epsfig}
\ifpdf
\usepackage[%
  pdftitle={Instructions for use of the document class
    elsart},%
  pdfauthor={Simon Pepping},%
  pdfsubject={The preprint document class elsart},%
  pdfkeywords={instructions for use, elsart, document class},%
  pdfstartview=FitH,%
  bookmarks=true,%
  bookmarksopen=true,%
  breaklinks=true,%
  colorlinks=true,%
  linkcolor=blue,anchorcolor=blue,%
  citecolor=blue,filecolor=blue,%
  menucolor=blue,pagecolor=blue,%
  urlcolor=blue]{hyperref}
\else
\usepackage[%
  breaklinks=true,%
  colorlinks=true,%
  linkcolor=blue,anchorcolor=blue,%
  citecolor=blue,filecolor=blue,%
  menucolor=blue,pagecolor=blue,%
  urlcolor=blue]{hyperref}
\fi

\makeatletter
\def\elsartstyle{%
    \def\normalsize{\@setfontsize\normalsize\@xiipt{14.5}}
    \def\small{\@setfontsize\small\@xipt{13.6}}
    \let\footnotesize=\small
    \def\large{\@setfontsize\large\@xivpt{18}}
    \def\Large{\@setfontsize\Large\@xviipt{22}}
    \skip\@mpfootins = 18\p@ \@plus 2\p@
    \normalsize
}
\@ifundefined{square}{}{}
\makeatother

\def\file#1{\texttt{#1}}

\begin{document}

\begin{frontmatter}
\title{A method based on transverse energy balance of jets for selection of 
direct photons and fragmentation photons in high energy pp collisions }

\author{Mriganka Mouli Mondal, Subhasis Chattophadyay}
\address{Variable Energy Cyclotron Centre, 1/AF Bidhan-nagar,
 Kolkata- 700064, India}

\ead{mmondal@veccal.ernet.in, sub@veccal.ernet.in}
%\ead[url]{}

\begin{abstract}
Direct photons are  important probes in high energy collisions. 
They play an important role in determining the parton distribution function 
directly
inside a proton as well as the nature of the matter formed 
in heavy ion collisions. However fragmentation photons play the role of 
prominent background in identifying the direct photons. In the present 
work we developed  a new 
method based on the transverse energy balance of jets for enrichment of 
direct photon
candidates in pp collisions. This method 
can reject  35$\%$ of the background photons (fragmentation) which can not be 
suppressed by isolation.  Efficiency of detection of direct photon decrease 
by 10$\%$
in the method.

\end{abstract}

\end{frontmatter}

\section{Introduction}
%\label{intro}
Direct photons in p-p collisions are very important in determining the parton distribution
function inside a nucleus. Since it gives direct control to the
parton kinematics, precise measurement of intrinsic transverse momentum
of partons can be made. Measurements of yields of direct photons can also be
used to compare with the pQCD predictions.
When one considers the detection of direct photons, the task at high 
energies e.g, RHIC and LHC  become quite difficult due to the 
presence of other sources of photons, acting as background to the direct 
photons.
\\
Direct photons, originated from two basic interactions e.g, 
$q \bar{q}\rightarrow g \gamma$
 and $qg \rightarrow q \gamma$ are very important observables in
 high energy collisions~\cite{one}.
Main sources of background photons are from hadronic decays
 and
 the$~~~$ fragmentation of partons. Some $~~$of the processes giving fragmentation 
photons$~~$ are qg $\rightarrow$ g(q$\gamma$) and qg $\rightarrow$ q(g$\gamma$)
% ( from NLO photon calculation $\alpha_{s}\alpha$, photons fron radiations  )
,  even  dominating in higher order  due to collinear singularity. It has
 been shown that the yields of these
photons are significant at relatively large $\sqrt{s}$. e.g, 
at $\sqrt{s} = 540$ GeV~\cite{A}\cite{B},
at RHIC energies and at LHC energies~\cite{C}\cite{D}.
However, theoretically  there are rather large uncertainty in the estimation 
of the yield
of fragmentation photons~\cite{E}, thereby making the measurement of 
fragmentation 
photons at pp collisions very important. This will enable to set a reference
for AA collisions, where direct photons can be used to calibrate the jets
and thereby allowing more precise analysis of jet quenching. Recently at
RHIC, efforts are made to measure fragmentation photons~\cite{F}.\\
For the measurement of direct photons, various methods (e.g. shower 
shape) are employed to identify and estimate decay photons (e.g.
 for $\pi^{0}$, $\eta$).
%\begin{figure}[h]
%\vbox{\hbox to\hsize{\psfig{figure=aaa.eps, scale=0.35,angle=0,  
%\vbox{\hbox to\hsize{\psfig{figure=six.eps,  scale=0.2, angle=0,  height=7.5cm, width=10.0cm}\hfill}}
%\caption{Rejection efficiency of fragmentation photons as a function of the fraction of jet energy carried by the fragmentation photon.}
%\end{figure}   
 Apart from these, the method being used
extensively for identifying the direct
photons is by the use of isolation cuts, since unlike direct photons,
the fragmented photons carry charge
particles as a part of jet. Hence the isolation cut is quite effective in
rejecting the decay and fragmentation photons for most of the cases. But 
this  cut is not
very effective  in 
rejecting fragmentation photons when they share most of the parton's energy
and appear like direct photons. It has been demonstrated extensively for the
case of ALICE experiment~\cite{G} that all the methods mentioned earlier will 
enrich the photon sample. However no special effort is made for the 
rejection of fragmentation photons.
It has been found that the fragmentation photons at LHC energies can be removed
at best by 55 $\%$ by the method of isolation.\\ 
\begin{figure}[b]
\vbox{\hbox to\hsize{\psfig{figure=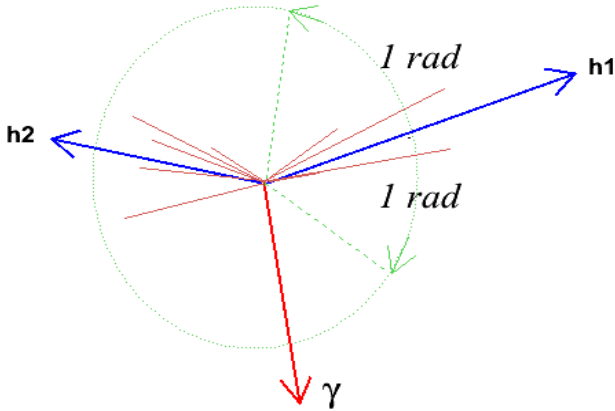, scale=0.37,angle=0,  height=6.5cm,
 width=10cm}\hfill}}
\caption{Event structure with fragmentation photons which escape isolation cut.}
\end{figure}
In this letter we are  presenting  a technique based on the transverse energy
of the away side
jets relative to the detected photon for discrimination of direct and
fragmentation photons. For fragmentation photons,
the  ratio of the
photons transverse energy to  the jet transverse energy likely to differ
 from the
ratio obtained for direct photons. In the present  method this property is
used
for the enrichment of photon samples.\\
The paper is organized as follows, in section-2, we discuss the methods in 
detail, simulation procedure, data set used and results are discussed in 
section-3. In section-4 we discuss the applicability of the method. 
\section{Method based on transverse energy balance of jets}
%\label{intro}
Direct photons originate from two basic processes:  (i) annihilation
of quarks f$\bar{f} \rightarrow$ g$\gamma$ and by (ii) Compton scattering
fg $\rightarrow$ f$\gamma$ (where f may be u, d, c, s, t, b and their
anti-pairs). On the other hand 
fragmentation photons  are originated 
by the process of fragmentation of partons into colorless particles
(iii)$f_{1}$f $\rightarrow$ ($f_{1} \gamma_{frag}$)f, where $\gamma_{frag}$ is 
the fragmentation photon originating from the parton $f_{1}$. \\
For the cases (i) and (ii), the transverse energy of the photon is equal to that
of the other parton($f/g$), i.e, $E_{T,\gamma_{dir}}
 = E_{T,f/g}$. In contrary, for the case (iii), the transverse 
energy of  photon is
always less than that of the parton f, i.e,  $E_{T,\gamma_{frag}} < E_{T,f}$.
From the experiments we do not 
get the partons but the jets of particles fragmented from the parton.
Hence in present method we have calculated $E_{T}$ of jet by summing over
all charged particles around the leading hadron within a specific phase space
in $\eta$ and $\phi$. After studying the jet topology on transverse plane, we
 have
found that jets fragments, on away side may not have unidirectional structure 
in a cone,
 it 
might even be visualized as two jets as shown in Fig.1. 
It should be noted that, as this is obtained from a
 limited phase space, calculated $E_{T}$  here might not represent the total
jet energy, but the parameter based on this  $E_{T}$, discussed later
has the property of discrimination of direct and fragmentation photons. 
In case of pp collisions, it is expected that the background underlying 
the jet is not significant compared to heavy ion collisions.\\
For this work, we have used PYTHIA (6.214) as event generator. Current study
 is performed
for the LHC energy (pp at $\sqrt{s}=$14 TeV). We have generated direct and
fragmentation  photon
samples in the $p_{T}$ ranges  $\>$10 - 500 GeV. Two sources of photons
are studied
separately keeping the spectra as obtained from the event generator. No
weightings are done for relative contribution of two sources of photons.
We have taken a coverage of photon detector as
$|\eta|\ge$ 0.12 and 220$\le\phi\le$340,  and charge particle detector as
 $|\eta|\le$ 1 with full azimuthal coverage.
 This coverage
 is used for the detectors in ALICE for the detection of photons and charge
particles~\cite{H}. In the definition of away-side regions we have taken 
two leading
hadrons h1, h2 (Fig.1) such that $|$$\phi_{h1} - \phi_{h2}$$|$ $>$ 1 rad.
 We calculated summed transverse energies as follows:   
%$E_{T1}$ = $\displaystyle\sum_{i=\pi^{\pm}(\phi>\phi_{h1}-1rad)}^
%{\phi<\phi_{h1}+1rad}E_{T_{i}}$\\
%$E_{T2}$ = $\displaystyle\sum_{i=\pi^{\pm}(\phi>\phi_{h2}-1rad)}^
%{\phi<\phi_{h2}+1rad}E_{T_{i}}$  \\\\
%If $E_{T1}$ $>$
%$E_{T2}$, we used $E_{T}$ = $E_{T1}$, else we have assigned 
%$E_{T}$ = $E_{T2}$. We then extracted the discriminating parameter, 
%``f'' = $E_{T,\gamma}$/$E_{T}$.
%
\begin{eqnarray}
E_{T1} = \displaystyle\sum_{i=\pi^{\pm}(\phi>\phi_{h1}-1rad)}^
{\phi<\phi_{h1}+1rad}E_{T_{i}}
\\
E_{T2} = \displaystyle\sum_{i=\pi^{\pm}(\phi>\phi_{h2}-1rad)}^
{\phi<\phi_{h2}+1rad}E_{T_{i}} 
\end{eqnarray}
If $E_{T1}$ $>$
$E_{T2}$, we used $E_{T}$ = $E_{T1}$, else we have assigned 
$E_{T}$ = $E_{T2}$. We then extracted the discriminating parameter, 
`f' = $E_{T,\gamma}$/$E_{T}$.
It is expected that `f' $>$ 1 for direct
$\gamma$ and  `f' $<$ 1 for fragmentation photons. Fig.2 shows the 
distribution
of `f' for the direct photons and the fragmentation photons  clearly
 demonstrating the
possibilities of using this variable for discriminating two sources of photons.
 There are cases
where deviations occur from this expectation because it might not be
possible to collect significant amount of away-side jet energy since the
jet
energy may spread beyond $|\eta|\le$ 1 and 
background from the underlying events might play a role. For the results 
presented here, we use `f' as the parameter for enrichment of direct
photon samples. 
It should be noted that by increasing the coverage of photon detectors, 
we can increase the statistics, but discriminating power will not change
significantly.\\
\begin{figure}[h]
  \vbox{\hbox to\hsize{\psfig{figure=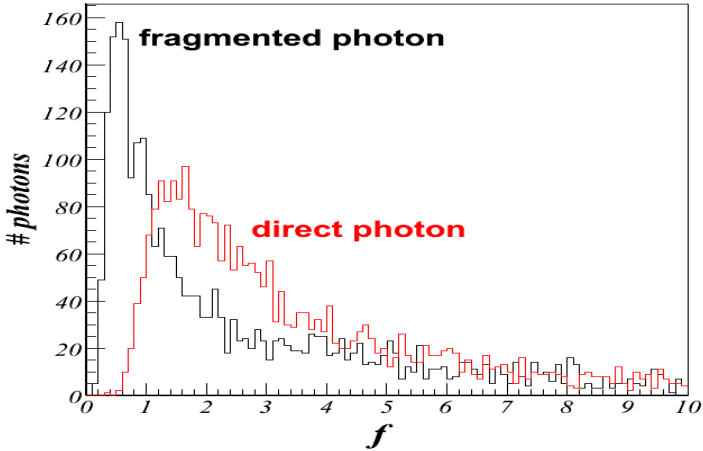, scale=0.37,angle=0,  height=6.5cm,
	width=10cm}\hfill}}
  \caption{Distribution of `f' for direct and fragmentation photons.}
\end{figure}
\section{Simulation and results}
\label{intro}
\begin{figure}[b]
  \vbox{\hbox to\hsize{\psfig{figure=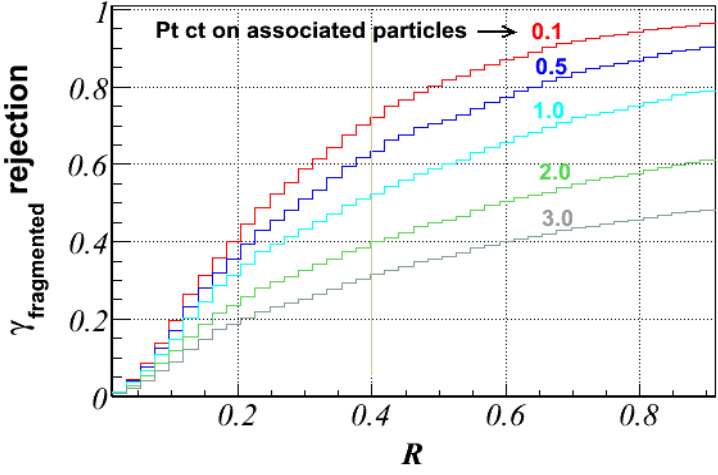, scale=0.37,angle=0,
	height=6.5cm, width=10cm}\hfill}}
  \caption{Fraction of fragmentation photons rejected after isolation cut.}
\end{figure}
\begin{figure}[b]
  \vbox{\hbox to\hsize{\psfig{figure=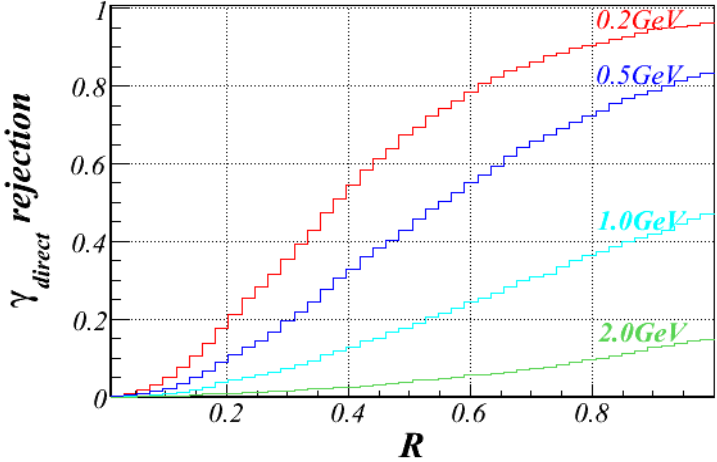, scale=0.37,angle=0,
	height=6.5cm, width=10cm}\hfill}}
  \caption{Fraction of direct photons rejected after isolation cut.}
\end{figure}
\begin{figure}[b]
  \vbox{\hbox to\hsize{\psfig{figure=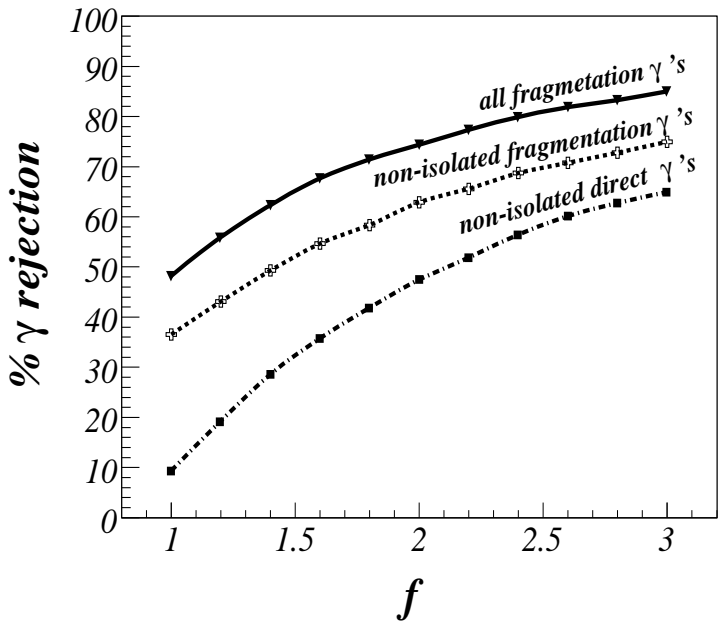, scale=0.3,angle=0,  
	height=8.5cm, width=10.1cm}\hfill}}
\caption{Fraction of non-isolated fragmentation (open cross markers)  and direct (filled rectangular markers)  photons rejected with $p_{T, associated}$ $>$ 1 GeV  and for various `f' values after isolation cut. Solid curve shows the rejection fraction for incident fragmentation photons with `f-cut' alone.}
\end{figure}     
In current approach we present the fractions of direct and fragmentation
photons rejected 
after the  application of  isolation cut and f-cut,  first separately and
then both applied sequentially.
Simulated data samples consists of  direct photons ($\gamma -$ jet) events,  
fragmentation
photons from jet-jet in the detector coverage mentioned earlier. 
Other set of criteria e.g. shower shape,  are not studied here as it 
depends strongly on detector properties. It is expected that, these will 
enrich the photon samples even further. \\
$~~~$Events are passed through varying isolation radii,
R = $\sqrt{(\Delta\eta)^{2}+(\Delta\phi)^{2}}$ with various thresholds in 
transverse momentum ($p_{T}$) on associated particles, where $\Delta\eta$ 
and $\Delta\phi$
are the separation of charged particles from the triggered photon
in $\eta$ and $\phi$ respectively. The results of isolation alone 
for fragmentation 
and direct photons  are shown in Fig.3 and in Fig.4 
respectively. We have plotted the fraction of fragmented and direct photons
 rejected
for different isolation radii and for different $p_{T}$ thresholds on
associated particles.
 It is seen that for $\Delta$R=0.4 and 
$p_{T, threshold}$ (associated particles) = 1 GeV, $\sim$55$\%$ of the 
fragmentation
photons are  rejected and  $\sim$10$\%$ of the direct photons are lost due
to the isolation criteria.\\
The results for the application of `f' cut alone is shown in fig.5 by 
solid line. The dashed lines show the result on  `f' cuts on the
photons which can not be rejected by isolation cuts. The percentage of
rejection for both the cases with the variation of `f' values are as 
follows $:$\\
1. The solid curve shows that only `f' cut alone can reject fragmentation 
photons in the range of 50-80$\%$.\\
2. The curve with filled points shows the direct photon rejection
vary from 10-60$\%$.\\
3. The curve with open cross marks shows the rejection efficiency of the 
fragmentation photons
which can not be removed by isolation cuts. We have applied `f' cuts for
R=0.4 and $p_{T, associated}$ $>$ 1 GeV. As shown in the dashed curve 
(cross markers) upto 70$\%$  of non-isolated fragmentation photon can be 
rejected by this cut.\\
The rejection of direct photons is always less compared to that of the 
the fragmentation
photons. Above some `f' value, two  cases become parallel, suggesting 
little increase in purity of direct photon samples, while not rejecting direct
photons appreciably.
 For f = 1 cut, $\sim$35$\%$ of the non-isolated fragmentation
photons can be removed in the new approach at the cost of 10$\%$ loss of
the direct photons.\\
The performance of `f' cut is dependent on 
the fraction of jet energy carried by the fragmentation photon. 
The study has been made  at $\sqrt{s}=$ 200 GeV with $p_{T}$ of fragmentation
 photons
above 4 GeV and jet $p_{T}$ $\ge$ 10 GeV. In Fig.6 we show the variation
of rejection efficiency in isolation method and f-cut method with
varying fraction of jet energy carried by the fragmentation photon. 
 It is clear from Fig.6 that for the fragmentation
 photons carrying larger fraction of  jet energy, identification
efficiency decreases and any of the methods perform similarly for direct and 
fragmentation photons. 
\begin{figure}[h]
%\vbox{\hbox to\hsize{\psfig{figure=aaa.eps, scale=0.35,angle=0,  
\vbox{\hbox to\hsize{\psfig{figure=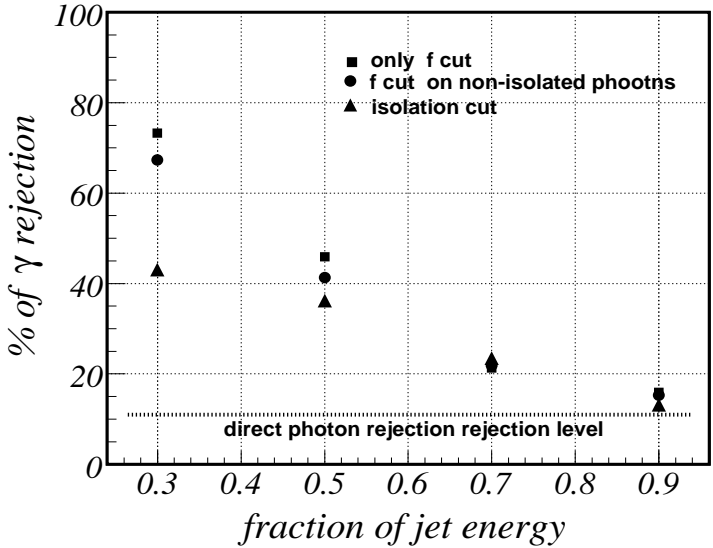,  scale=0.2, angle=0,  height=7.5cm, width=10.0cm}\hfill}}
\caption{Rejection efficiency of fragmentation photons as a function of the fraction of jet energy carried by the fragmentation photon.}
\end{figure}   
 \section{Summary and Discussions}
%\label{intro}
Based on the differences in jet topology for direct photons and fragmentation 
photons, we have developed a method to enrich direct photon sample further to 
the result obtained for isolation cut. The method is based on the transverse 
energy balance of near and away side jets for two sources of photons. 
If we consider the role of the fragmentation photon in deciding the purity
of direct photon detection assuming that the decay photons are completely
rejected, the ratio of yields of direct to fragmentation  plays
an important role. At LHC energy, for pp collisions, if we take the ratio as 1  in the $p_{T}$
region we are looking for  we obtain $\sim$ 50$\%$ fragmentation photons
rejected due to isolation cut and 40$\%$ non-isolated fragmented photon 
rejected 
by `f' cut at the cost of 10$\%$ direct photon loss. We can estimate
 that the overall
purity of direct photon increases from 66$\%$ to 75$\%$.


\begin{thebibliography}{99}
 
\bibitem{one} J.~F.~Owens, 
% Large-momentum-transfer production of direct photons, jets, and particles
Rev. Mod. Phys. {\bf 59}, 465 (1987).
\bibitem{A} C.~Albajar {\it et al} [by UA1 Collaboration], Phys. Lett. B {\bf 209}, 385 (1988).
% Direct Photon Production at the CERN Proton - anti-Proton Collider. 
\bibitem{B} R.~Gandhi, F.~Halzen and F.~Herzog, 
% Direct Photons In Jets
Phys. Lett. B. {\bf 152} 261 (1984).
\bibitem{C} P.~Aurenche, R.~Baier, M.~Fontannaz,
% Prompt photon production at colliders 
 Phys. Rev. D {\bf 42}, 1440 (1990).
\bibitem{D} P.~Aurenche, P.~Chiappetta, M.~Fontannaz, J.~.Ph.~Guillet, E.~Pilon
%Next to leading order bremsstrahlung contribution to prompt-phootn production 
 Nucl. Phys. B {\bf 399}, 34 (1993).
%\bibitem{E} M.~Gl$\ddot{u}$ck, L.~E.~Gordon, E.~Reya, and W.~Vogelsang,
\bibitem{E} M.~Gl\"uck, L.~E.~Gordon, E.~Reya, and W.~Vogelsang,
%High-pt Phootn Production at pp(bar) colliders
 Phys Rev. Lett. {\bf 73}, 388(1994).
\bibitem{F} D.~Peressounko {\it et al} [for PHENIX collaboration],
%Direct Photon production in p+p and d+Au collisions measured with PHENIX Experiment
 Nucl. Phys. A {\bf 783}, 577c (2007).
\bibitem{G} ALICE Collaboration {\it et al}, J. Phys. G: Nucl. Part. Phys. {\bf 32}, 
1295 (2006).
\bibitem{H} ALICE Collaboration {\it et al}, J. Phys. G: Nucl. Part. Phys. {\bf 30}, 
1517 (2004).

\end{thebibliography}
\end{document}